\documentclass[prl,twocolumn,showpacs,superscriptaddress]{revtex4}

\usepackage{ulem}
\usepackage{amsmath}
\usepackage{graphicx,color}
\usepackage{dcolumn}
\usepackage{bm}
\begin{document}
\title{Spin squeezing of a cold atomic ensemble with the nuclear spin of one-half}

\affiliation{Department of Physics, Graduate School of Science, Kyoto University, Kyoto 606-8502, Japan}
\author{T. Takano}
\affiliation{Department of Physics, Graduate School of Science, Kyoto University, Kyoto 606-8502, Japan}
\author{M. Fuyama}
\affiliation{Department of Physics, Graduate School of Science, Kyoto University, Kyoto 606-8502, Japan}
\author{R. Namiki}
\affiliation{Department of Physics, Graduate School of Science, Kyoto University, Kyoto 606-8502, Japan}
\author{Y. Takahashi}
\affiliation{Department of Physics, Graduate School of Science, Kyoto University, Kyoto 606-8502, Japan}
\affiliation{CREST, JST, 4-1-8 Honcho Kawaguchi, Saitama 332-0012, Japan}

\date{\today}

\begin{abstract}
To establish an applicable system for advanced quantum information processing between light and atoms, we have demonstrated the quantum non-demolition (QND) measurement with a collective spin of cold ytterbium atoms ($^{171}\mathrm{Yb}$), and observed $\mathrm{1.8 ^{+2.4}_{-1.5}}$ $\mathrm{dB}$ spin squeezing. Since a $^{171}\mathrm{Yb}$ atom 
has only a nuclear spin of 1/2 in the ground state, the system is the simplest spin ensemble and robust against decoherence. 
We used very short pulses with the width of 100 $\mathrm{ns}$, so the interaction time became much shorter than the decoherence time,
which is important for multi-step quantum information processing.
\end{abstract}
\maketitle
\parskip=0pt
Quantum non-demolition (QND) measurements are measurements in which the strategy is chosen to evade the undesirable
effect of back action \cite{QND-review,LightQND}.
They have been developed to manage the quantum noise, and are also useful for a quantum-state preparation device and producing a quantum entanglement \cite{EPR} as well as a feasible model to capture basic features of a quantum measurement process \cite{QND-review,LightQND}. 
Previously, the QND measurement of the photon number and the amplitude quadrature of light have been realized \cite{double,quadrature,photonnumber,Haroche}. 
The QND measurement of the collective spin is also considerably interesting, and 
in fact the QND interaction of collective spin of an atomic ensemble (spin-QND interaction)
via the Faraday-rotation interaction with linearly-polarized off-resonant light has been proposed \cite{QND,Kuzmich}. An implication is the spin squeezed state \cite{Kitagawa-Ueda}, which could improve the measurement precision of the atomic clock transition \cite{Clock,Noise} and of the permanent electric dipole moment to test the violation of time reversal symmetry \cite{Flambaum}. The spin-QND interaction is also useful for
implementing continuous-variable quantum information devices, such as quantum memory and quantum teleportation \cite{EPR,tele,memoryExp,retr,Takano}. The variety of the interactions and tunability of their strength are useful characteristics  of atoms whereas the property of the interaction is rather fixed by the parameter of the non-linear crystal for the case of the QND measurement in optics \cite{QND-review}. 

 Previous experimental approaches for the spin-QND interaction \cite{EPR,Bigelow} used thermal alkali atoms and continuous-wave light or long pulsed light of typically 1 $\mathrm{ms}$ width, which is comparable with the decoherence time of the atomic spin \cite{EPR,Bigelow}. Hence, it is essential to implement the interaction with shorter pulses and more controllable cold atoms for composing the quantum interface where more-than-twice interactions between the atoms and light beam are required \cite{Takano}. In addition, it should be noted that the description of the spin-QND interaction is based on the standard model of the collective spin composed of the spin one-half atoms \cite{Kupriyanov,Stockton}. 
However, the cesium atoms used in the previous experiments
have more complicated multi-level structures, which causes serious
difficulties
as is pointed out in Ref. \cite{Kupriyanov,Stockton}.
Therefore, it is widely valuable to demonstrate the spin-QND interaction with cold spin one-half atoms and short pulses.
 
In this Letter, we report the successful experimental realization of the spin-QND measurement with laser-cooled ytterbium ($^{171}\mathrm{Yb}$) atoms and short light pulses.
The $^{171}\mathrm{Yb}$ atom has the simplest ground state with a nuclear spin of one-half and has no electron spin.
The system will be robust against a stray magnetic field because the magnetic moment of a nuclear spin is thousandth of that of an electron spin. 
By using a short light pulse with the width of 100 $\mathrm{ns}$, more-than-hundred-time operations are expected to be achieved within the coherence time.
To show the realization of the spin-QND interaction, we have investigated the correlations between the two light pulses which sequentially interact with the atoms. 
\par
We note that the spin-QND measurement for $^{171}\mathrm{Yb}$ atoms realized in this work is especially important since
the cold $^{171}\mathrm{Yb}$ atoms in an optical lattice is considered to be one of the promising candidates for the
future optical standard \cite{Katori,Yb-OLC}.
The quantum noise limited performance of the $^{171}\mathrm{Yb}$-based optical lattice clock would be significantly improved by mapping the
ground-state spin squeezing onto the clock transition $^1\mathrm{S}_0\leftrightarrow {^3\mathrm{P}_0}$ \cite{Noise}.
In this direction, we have recently learned that the spin squeezing for the clock transition between the hyperfine states of alkali-atoms
is realized by a somewhat different method \cite{Vuletic}.

 To describe the Faraday-rotation interaction, let us define the normalized collective spin operator of the atoms $\vec{{\tilde{J}}} =(\tilde{J_x},\tilde{J_y},\tilde{J_z})=(1/\sqrt{N_A/2})\sum_{i=1}^{N_A} \vec j_i$, where $\vec{j}_i$ is a spin operator of a single atom and $N_A$ is the number of the atoms. 
The normalized Stokes operator of a pulsed light $\vec {\tilde{S}} =(\tilde {S_x},\tilde {S_y},\tilde {S_z})$
is defined by $\tilde{S_x} = (2\sqrt{N_L})^{-1}\int _0^t(a_+^{\dagger}a_-+a_-^{\dagger}a_+)dT$, 
$\tilde{S_y} = (2i\sqrt{N_L})^{-1}\int _0^t(a_+^{\dagger}a_--a_-^{\dagger}a_+)dT$, $\tilde{S_z} = (2\sqrt{N_L})^{-1}\int _0^t(a_+^{\dagger}a_+-a_-^{\dagger}a_-)dT$, where $N_L$ is the mean photon number of the pulse, 
$t$ is the pulse duration, and $a_{\pm}$ is the
annihilation operator of $\sigma _{\pm}$ circular polarization mode, respectively \cite{Duan}.
In our experiment, we consider the situation that the initial states of the light and atoms are coherent states and polarized in the $x$-direction, namely, $\tilde{J_x}\simeq \sqrt{J}\equiv \sqrt{N_A/2}$ and $\tilde{S_x}\simeq \sqrt{S}\equiv \sqrt{N_L/2}$ hold. 
Then, the angular momentum commutation relation for each of $\vec {\tilde {J}}$ and $\vec {\tilde{S}}$ 
implies $[\tilde{J_y}, \tilde{J_z}]=i$ and $[\tilde{S_y},\tilde{S_z}]=i$,
and they satisfy the uncertainty relation,  $\langle\Delta \tilde{J^2_y}\rangle\langle\Delta \tilde{J^2_z}\rangle\geq  1/4$ and $\langle\Delta \tilde{S^2_y}\rangle\langle\Delta \tilde{S^2_z}\rangle\geq  1/4$.
The variances of coherent states are $\langle\Delta\tilde{S^2_y}\rangle=\langle\Delta\tilde{S^2_z}\rangle=1/2$ and $\langle\Delta \tilde{J^2_y}\rangle\!=\!\langle\Delta\tilde{J^2_z}\rangle=1/2$. We say the state is spin-squeezed in the $z$ direction if $\langle\Delta\tilde{J^2_z}\rangle< 1/2$ is satisfied \cite{Kitagawa-Ueda}. 

The Hamiltonian of the Faraday-rotation interaction is given by
\begin{equation}
H_{int}=\alpha \sqrt{SJ}\tilde{S_z}\tilde{J_z},
\end{equation}
where $\alpha$ is a real constant and $z$ means the propagation direction of the light \cite{QND}.
This interaction causes the time evolution of $\vec {\tilde{J}}$ and $\vec {\tilde{S}}$, 
so that, 
$
\tilde{S_y}\to\tilde{S_y} + \kappa\tilde {J_z}$, $\tilde{S_z}\to\tilde{S_z}$,
$\tilde{J_y}\to\tilde{J_y} + \kappa \tilde{S_z}$, $\tilde{J_z}\to\tilde{J_z}
$, where the interaction strength is given by $\kappa \equiv \alpha t\sqrt{JS}$.
Note that this interaction satisfies a back-action evading condition $[H_{int}, \tilde{J_z}]\!\!=\!\!0$,
 and makes a quantum correlation between $\tilde {J_z}$ and $\tilde{S_y}$. Thereby, the measurement of $\tilde{S_y}$ will essentially project the spin state into an eigenstate of $\tilde {J_z}$ and the variance of $\tilde {J_z}$ will be squeezed.
The measurement of $\tilde {S_y}$ is said to be the QND measurement of $\tilde {J_z}$,
and induces the spin squeezing in the $z$ direction \cite{QND}.

Suppose that two successive pulses interact with the atoms. Then, the Stokes operator of the first pulse $
(\tilde{S}_{1,y},\tilde{S}_{1,z})$ and that of the second pulse $
(\tilde{S}_{2,y},\tilde{S}_{2,z})$ are transformed as
\begin{align}
&\tilde{S}_{1,y}^{(f)}=\tilde{S}_{1,y}^{(i)} + \kappa \tilde {J}_{1,z}, &\tilde{S}_{1,z}^{(f)}=\tilde{S}_{1,z}^{(i)}, \notag\\
&\tilde{S}_{2,y}^{(f)}=\tilde{S}_{2,y}^{(i)} + \kappa \tilde {J}_{2,z}, &\tilde{S}_{2,z}^{(f)}=\tilde{S}_{2,z}^{(i)} \label{time},
\end{align}
 where $\tilde{J}_{1,z}$ and $\tilde{J}_{2,z}$ are the collective spin operators for the first pulse and second pulse, respectively. If both of the initial states are coherent states and the QND condition $\tilde{J}_{1,z}=\tilde{J}_{2,z}$ is satisfied, 
we have 
\begin{align}
&\sigma_{1}\equiv \langle \Delta (\tilde{S}_{1,y}^{(f)})^2 \rangle = (1+\kappa^2)/2 \\
&\sigma_{2}\equiv \langle \Delta (\tilde{S}_{2,y}^{(f)})^2 \rangle = (1+\kappa^2)/2 \\
&\sigma_{z}\equiv \langle \Delta (\tilde{S}_{1,z}^{(f)})^2 \rangle = \langle \Delta (\tilde{S}_{2,z}^{(f)})^2 \rangle = 1/2.
\end{align}
This implies that the spin-QND interaction increases the individual variances $\sigma_{1,2}$ by the same factor of $\kappa ^2 /2$ and it does not change  the variances in the $z$ direction, $\sigma_z$.
In addition, we have the following relations about the positive correlation $\sigma _{+}$ and the negative correlation $\sigma_{-}$ as
\begin{align}
&\sigma_{+}\equiv \langle \Delta (\tilde{S}_{1,y}^{(f)} + \tilde{S}_{2,y}^{(f)}) ^2\rangle/2=(1+2\kappa^2)/2\\
&\sigma_{-}\equiv \langle \Delta (\tilde{S}_{1,y}^{(f)} - \tilde{S}_{2,y}^{(f)}) ^2\rangle/2=1/2.
\end{align}
As one can see, $\sigma _{+}$ increases by a factor of $\kappa ^2$, while $\sigma _-$ does not change. 
In contrast with the QND measurement in optics \cite{LightQND,double}, it is difficult to directly  measure the spin states, and hence, we experimentally investigate these relations to confirm the achievement of the spin-QND measurement. 
Specifically, we experimentally show the behavior of the variances and correlations for various incident photon number of the pulses $N_L$.
In order to investigate the behavior when the QND condition $\tilde{J}_{1,z}=\tilde{J}_{2,z}$ does not hold,
we perform the same measurement with the atomic spins re-initialized during the interval of the two interactions. 
In this case, all of the variances are expected to be $(1+\kappa ^2)/2$.
We also estimate the measurement-conditioned variance and the degree of the spin squeezing obtained by the experiments.
\par
\begin{center}
\begin{figure}[h]
\includegraphics[width=8cm]{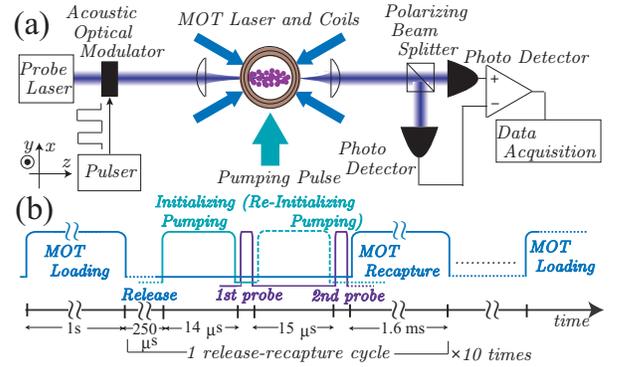}
\caption{(color online) (a)Experimental setup.
Two successive linearly-polarized 100 $\mathrm{ns}$ probe pulses pass through the polarized $^{171}\mathrm{Yb}$ atoms released from MOT 
and the polarizations of the pulses are measured.
(b)Experimental time sequence. 
At first, typically $10^6$ atoms are loaded in MOT and are released in the next period, during
which the trapping system is switched off.
Secondly, the atoms are polarized by the circularly-polarized resonant pumping pulse.
Then, the two probe pulses pass through the atoms and go into the polarization detector. 
The atoms can be re-initialized when necessary, by applying the optical pumping pulse between the two pulses, represented by the dashed line.  
}
\end{figure}
\end{center}
\par
The experimental setup is shown in Fig. 1(a) and is basically the same as our previous experiment \cite{Fara}.
By several improvements, the Faraday-rotation angle of $\phi = \alpha t J/2 \simeq 143$ $\mathrm{mrad}$ was achieved with the relatively small 
fluctuation $\sigma (\phi )\simeq 10$ $\mathrm{mrad}$, where 
$\sigma(X)$ means the standard variance of $X$.
For the atomic number, we typically have $J\simeq 3.4\times 10^5$ and $\sigma(J)\simeq 2.4 \times 10^4$.
The probe system was the same as our another experiment \cite{pola}.
The probe beam was focused with the beam waist of $w _0=58$ $\mathrm{\mu m}$ and
the frequency of the light was locked to the center of the two hyperfine-splitted optical lines of the $^1\mathrm{S}_0\leftrightarrow{^1\mathrm{P}_1}$ transitions of $^{171}\mathrm{Yb}$.
 $\kappa$ is calculated as
\begin{equation}
\kappa = \frac{\Gamma \sigma_0 \sqrt{SJ}}{3\pi  w_0 ^2}\big( \frac{\delta -\delta _0}{(\delta -\delta _0)^2+ (\Gamma /2)^2}-\frac{\delta}{\delta ^2+ (\Gamma /2)^2} \big),
\end{equation}
where $ \Gamma = 2\pi \times  29$ $\mathrm{MHz}$ is the natural full line width, $\sigma_0 \!\!=\!\! 7.6\times 10^{-14}$ $\mathrm{m^2}$ is the photon-absorption cross section of the $^{171}\mathrm{Yb}$ atom, $\delta = 2\pi \times 160$ $\mathrm{MHz} $ is the 
detuning from the $^1\mathrm{S}_0\leftrightarrow {^1\mathrm{P}_1}$ ($F'= 3/2$) states, and
$ \delta _0\!\! =\!\! 2\pi \times 320$ $\mathrm{MHz}$ is the frequency difference between the $F'=1/2$ and $F'=3/2$ states in the $^1\mathrm{P}_1$ state \cite{Duan,Fara}.
In our experimental conditions, the maximum value of $\kappa$ is 0.62.
At this value, $N_L$ is $3.2 \times 10^{6}$ and
 the loss parameter $\epsilon \equiv rt/2$ is $9.3 \times 10^{-2}$, where $r$ is the absorption rate \cite{this}. 
\par
The time sequence is shown in Fig. 1(b).
At first, typically $10^6$ atoms are loaded in a magneto-optical trap (MOT) in 1 $\mathrm{s}$ and are released in the next 250 $\mathrm{\mu s}$, during
which the MOT light, the MOT magnetic field, and the Zeeman Slower light are switched off.
Secondly, the atoms are polarized by the circularly-polarized resonant pumping pulse whose width is 14 $\mu s$.
Then, two linearly-polarized probe pulses pass through the atoms and go into the polarization detector. 
The pulses have the same width of 100 $\mathrm{ns}$ and the interval between them is 15 $\mathrm{\mu s}$. 
The atoms can be re-initialized when necessary, by applying the optical pumping pulse during this 15 $\mathrm {\mu s}$ period. 
This process is represented by the dashed line in Fig. 1(b).  
The detection of the second pulse completes the single measurement process of our experiment. Then, the atoms are recaptured by MOT and reused for the next cycle. For the single loading of atoms, we repeated 10 release-and-recapture cycles as shown in Fig. 1(b).
In this manner, we measures about 2600 pairs of the Stokes operators of the pulses for each arrangement of the experimental parameters \cite{zero-vias}.
\begin{center}
\begin{figure}[h]
\includegraphics[width=7.5cm]{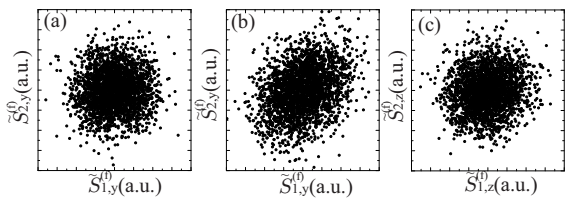}
\caption{Joint distribution of the measured polarization $\tilde {S}_{1,y}^{(f)}$ and $\tilde{S}_{2,y}^{(f)}$.
(a)No atoms. The variances were limited by the light shot-noise and the distribution was isotropic.
(b)With $8\times 10^5$ atoms in the probe region. 
The correlation was observed.
(c)$\tilde {S}_{1,z}$ and $\tilde{S}_{2,z}$ with $8\times 10^5$ atoms. The distribution is essentially the same as that of FIG. 2(a), as expected from Eq.(5).
}
\end{figure}
\end{center}
\par
Figure 2(a-b) show the joint distribution of the measured $\tilde {S}_{1,y}^{(f)}$ and $\tilde{S}_{2,y}^{(f)}$ for the cases with (a) no atoms and
(b)$8\times 10^5$ atoms at $\kappa = 0.62$ in the QND condition, respectively.
We observe the increase of $\sigma _{1,2}$ in (b) from the ones in (a) and the positive correlation in (b).
Figure 2(c) shows the measured joint distribution of $\tilde {S}_{1,z}^{(f)}$ and $\tilde{S}_{2,z}^{(f)}$ at $\kappa = 0.62$. The measurement of $\tilde{S}_{1,z}^{(f)}$
 and $\tilde {S}_{2,z}^{(f)}$ was performed by inserting a $\lambda /4$ plate before the polarizing beam splitter.
We can see that $\tilde{S}_{1,z}^{(f)}$ and $\tilde{S}_{2,z}^{(f)}$ do not show any specific correlation and the distribution is essentially the same as that of Fig. 2(a), as expected from Eq. (5).
\begin{center}
\begin{figure}[h]
\includegraphics[width=7cm]{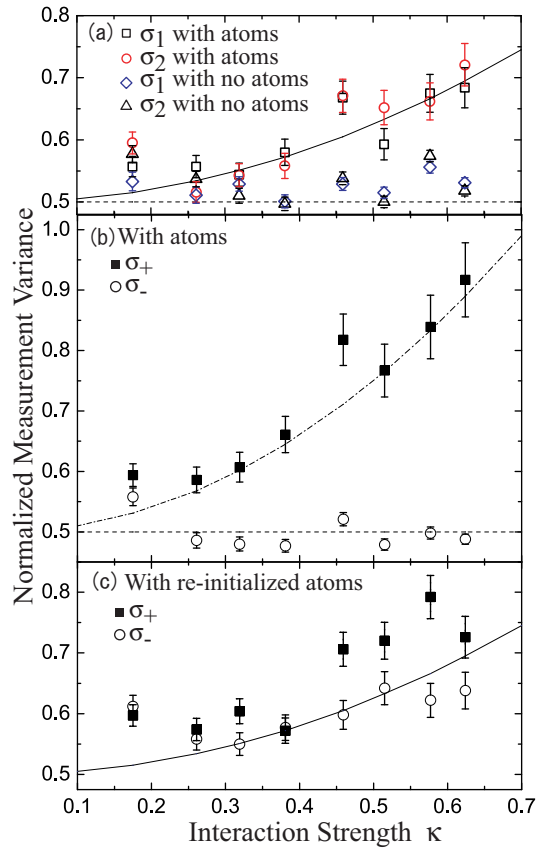}
\caption{(color online) Measured variance as a function of the interaction strength $\kappa$.
The solid curve, dash-dotted curve, and dashed curve are given by $(1+\kappa ^2)/2$, $(1+2\kappa ^2)/2$, and $1/2$, respectively.
(a)Individual variances $\sigma _{1,2}$ with and without atoms. 
At a large $\kappa$, it is apparent that $\sigma _{1,2}$ with atoms became larger than the ones without atoms.
(b)$\sigma_\pm$ with atoms.
As expected from Eq. (3-7), we observed larger variance of $\sigma_+$ than $\sigma _{1,2}$ with atoms in Fig. 3(a), 
while $\sigma_-$ remains the value of 1/2 corresponding to the light-shot noise.
(c)$\sigma_\pm$ with re-initialized atoms.
The significant difference between $\sigma_{\pm}$ observed in (b) almost disappeared.}
\end{figure}
\end{center}
\par
We have performed this measurement for a various interaction strength $\kappa$ by changing the incident photon number $N_L$ and observed the behavior of the variances and correlations. 
Figure 3(a) shows the individual variances $\sigma _{1,2}$ with and without atoms.
At a large $\kappa$, it is apparent that $\sigma _{1,2}$ with atoms became larger than the ones without atoms.
Figure 3(b) shows the values of $\sigma_\pm$ with atoms.
As expected from Eqs. (3-7), we successfully observed larger variance of $\sigma_+$ than $\sigma _{1,2}$ with atoms in Fig. 3(a), while $\sigma_-$ remains the value of 1/2
 corresponding to the light-shot noise.
Figure 3(c) shows the case with re-initialized atoms, where the above correlation observed in Fig. 3(b) almost disappeared and $\sigma _\pm$ had almost the same values as $\sigma _{1,2}$ in Fig. 3(a), as expected.
Here, the solid curve, dash-dotted curve, and dashed curve are given by $(1+\kappa ^2)/2$, $(1+2\kappa ^2)/2$, and $1/2$, respectively [see Eqs. (3-7)].
Note that the error bars are calculated from the statistical error of the variance measurement and the atomic number fluctuation.
As one can see, all of the measured variances are almost consistent with the theory.
\begin{center}
\begin{figure}[h]
\includegraphics[width=7cm]{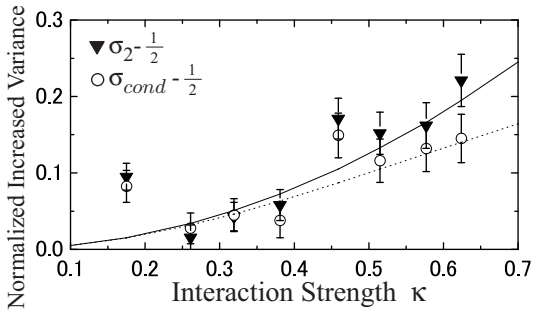}
\caption{Conditioned variance $\sigma _{cond} -1/2$ and $\sigma _2-1/2$ 
as a function of a various interaction strength $\kappa$.
Solid curve shows the theoretical value of the total variance $\kappa^2 /2$ and 
dotted curve shows the one of the conditioned variance, which is given as $\kappa^2/\{2(1+\kappa^2)\}$.
At a large $\kappa$, it is apparent that $\sigma _{cond}$ became smaller than $\sigma _{2}$.}
\end{figure}
\end{center}
\par
Finally, we show that the $z$-component of the spin $J_z$ is conditionally squeezed.
To observe this spin squeezing, one may count the data of $\tilde{S}_{2,y}^{(f)}$ only when $\tilde{S}_{1,y}^{(f)}$ takes a specific value. 
So, we divided the data around the center into 21 bins of $\tilde{S}_{1,y}^{(f)}$ and took the average of the variance 
of $\tilde{S}_{2,y}^{(f)}$ of each bin.
In Fig. 4, we showed the conditioned variance $\sigma _{cond}-1/2$ and $\sigma _2-1/2$ for various interaction strength $\kappa$.
At a large $\kappa$, it is apparent that $\sigma _{cond}$ became smaller than $\sigma _{2}$.
Here, the solid curve shows the theoretically expected dependence of the total variance $\kappa^2 /2$, and 
the dotted curve shows the theoretically expected one of the conditioned variance $\kappa^2/\{2(1+\kappa^2)\}$ \cite{QND}.
As one can see, the experimental results have the values near to the above theoretical estimation and 
the spin squeezing was achieved
with the degree of $\mathrm{1.8 ^{+2.4}_{-1.5}}$ $\mathrm{dB}$ when $\kappa = 0.62 $.

In conclusion, we reported the spin-QND measurement with cold $^{171}\mathrm{Yb}$.
From the quantitative and qualitative analysis, we concluded that we have achieved the spin-QND measurement and $\mathrm{1.8 ^{+2.4}_{-1.5}}$ $\mathrm{dB}$ spin squeezing.
This demonstration is widely valuable because we used cold atoms with the nuclear spin one-half and short pulses with the width of 100 $\mathrm{ns}$.
Important future step is multi-step quantum information processing \cite{Takano}, and the improvement of the $\mathrm{Yb}$ based optical lattice clock \cite{Yb-OLC}.

This work was supported by the Grant-in-Aid for Scientific Research of JSPS (Contracts No. 18043013 and
No. 18204035), SCOPE-S, and Global COE Program ``The Next Generation of Physics, Spun from Universality and Emergence'' from MEXT of Japan.
T. T. and R. N. are supported by JSPS Research Fellowships.


\begin{thebibliography}{99}
\bibitem{this}The effect of the loss is discussed in Ref. \cite{Duan}
\bibitem{zero-vias}Note that we have compensated a zero-vias of the measured data depending on which one of the 10 cycles the data belongs to,
but the behavior of the data is essentially the same when there is no compensation.
\end{thebibliography}
\end{document}